# High-Throughput Vector Similarity Search in Knowledge Graphs

Jason Mohoney*†, Anil Pacaci*, Shihabur Rahman Chowdhury*, Ali Mousavi*,
Ihab F. Ilyas*, Umar Farooq Minhas*, Jeffrey Pound*, Theodoros Rekatsinas*
*Apple, †University of Wisconsin-Madison

## ABSTRACT

There is an increasing adoption of machine learning for encoding data into vectors to serve online recommendation and search use cases. As a result, recent data management systems propose augmenting query processing with online vector similarity search. In this work, we explore vector similarity search in the context of Knowledge Graphs (KGs). Motivated by the tasks of finding related KG queries and entities for past KG query workloads, we focus on *hybrid vector similarity search (hybrid queries for short)* where part of the query corresponds to vector similarity search and part of the query corresponds to predicates over relational attributes associated with the underlying data vectors. For example, given past KG queries for a song entity, we want to construct new queries for new song entities whose vector representations are close to the vector representation of the entity in the past KG query. But entities in a KG also have non-vector attributes such as a song associated with an artist, a genre, and a release date. Therefore, suggested entities must also satisfy query predicates over non-vector attributes beyond a vector-based similarity predicate. While these tasks are central to KGs, our contributions are generally applicable to hybrid queries. In contrast to prior works that optimize online queries, we focus on enabling efficient batch processing of past hybrid query workloads. We present our system, HQI, for high-throughput batch processing of hybrid queries. We introduce a workload-aware vector data partitioning scheme to tailor the vector index layout to the given workload and describe a multi-query optimization technique to reduce the overhead of vector similarity computations. We evaluate our methods on industrial workloads and demonstrate that HQI yields a 31× improvement in throughput for finding related KG queries compared to existing hybrid query processing approaches.

## 1 INTRODUCTION

Vector similarity search has become a critical component in recommendation systems and search engines. Modern machine learning models are increasingly used to embed complex data such as images, text, or entities in knowledge graphs into vector representations that retain semantically meaningful information [5, 6, 9, 18, 29]. Similarity search over these vectors enables more accurate and contextualized search [10, 13, 15, 36] and recommendations [26, 27, 31, 33, 41] over multi-modal data. Due to the universality of vector embeddings, there is a recent increase in *vector database* offerings. A new breed of data management systems, such as ADBV [44], Milvus [42], Pinecone [2], Vearch [25], Vespa [3] and Vectara [1], augments query processing with vector similarity search primitives to power workloads where query processing requires searching over vectors to find the most similar ones to a query vector.

In this work, we study constrained vector similarity search for powering applications over industrial-scale Knowledge Graphs (KGs). Such applications include finding related entities, performing link prediction, and detecting erroneous facts [5, 19, 43]. These workloads require batch processing of *hybrid queries*, where a hybrid query consists of two parts: (i) vector similarity search and (ii) evaluation of predicates over relational attributes. For example, consider a service that recommends related artists given a song as input. Given a collection of millions of songs for which we want to support in such a service we may want to pre-compute all related—similar in a vector space—entities to each song that are of type Artist. Current approaches focus on online query processing and lack necessary optimizations to support high-throughput batch processing. To address this, we introduce a suite of system optimizations that enable high-throughput hybrid query processing over industrial-scale KGs. While we introduce these optimizations in the context of KG-related tasks, the optimizations are general and can be incorporated in general-purpose vector database systems.

*Motivating Workloads and Requirements.* We recently introduced Saga [19], a platform for constructing and serving knowledge at industrial scale. As part of this effort, we use *similarity search over KG embeddings* to solve tasks such as finding *related KG queries or entities* for past user queries and *link prediction* for missing fact imputation, among others [19].

For related KG queries, our goal is to construct and pre-evaluate a set of KG queries that are related to past user queries; the volume of queries served by our system offers significant opportunities for caching as queries are repeated across users. Such a service can power richer user experiences. For example, given the user query "How tall is Taylor Swift?" we want to construct queries of the form "How tall is person?" for people that are similar/related to Taylor Swift. For link prediction, our goal is to enrich the KG with new inferred facts (*e.g.*, imputing missing "collaborator" facts for "Taylor Swift") obtained via vector similarity search [47]. These use cases exhibit one or more of the following characteristics:

- **Hybrid queries:** We need to evaluate *hybrid queries* that combine vector similarity with relational attribute predicates. For the query "How tall is Taylor Swift?", we need to find the top entities that are close to "Taylor Swift" in the KG embedding vector space but they must also satisfy the predicates that their entity type is equal to "Person" and they have a non-NULL value for attribute "height". The last two predicates are necessary to get valid responses for the identified related queries. Similarly, imputing missing "collaborator" facts for "Taylor Swift" requires performing link prediction to identify entities of type "Artist" that are close to "Taylor Swift" conditioning on the predicate "collaborator".
- **Batch processing:** In contrast to online similarity search, these workloads often need to be processed in a batch setting, *e.g.*, find related queries for a batch of past user queries, or perform link prediction for all KG entities [19]. To guarantee efficient processing, we need a query evaluation design that prioritizes throughput similarly to analytical workloads.
- **Availability of prior workload:** We find that industrial question answering KG workloads exhibit *filter commonality* and *filter stability*, thus, allows us to tailor the vector database design to the workload characteristics [38]. Furthermore, the relational predicates present in a hybrid query are often correlated with the vectors used to compute similarity (*e.g.*, attribute "height" is correlated with entities representing people). Therefore, a *workload-aware approach* can help optimize vector search by accounting for these properties.

The above requirements necessitate new optimizations that current vector database systems do not cover. First, most systems provide limited support for relational attribute predicates i.e., primarily numerical comparisons, and they are not optimized for queries with multiple predicates. Second, all existing systems treat the relational predicates and vector similarity search as separate queries whose results are merged to generate final results. Doing so can lead to unnecessary computational overhead or low recall, as we later show in Section 6. Third, all existing systems aim to reduce the latency of individual online queries and do not optimize for throughput. However, many of our KG workloads evaluate hybrid queries in bulk, requiring high-throughput batch processing. Lastly, the presence of past query workloads provides opportunities to leverage the data and workload distributions for devising algorithms and data structures tailored to particular use cases [24, 38]. Custom fitting system components to a given workload and data has proven to provide significant performance improvements over their traditional counterparts in the context of relational workloads [4, 7, 17, 38, 46]. To the best of our knowledge, no vector database system utilizes *a priori* workload information for workload-aware database design.

Here we describe the key components of our proposed Hybrid Query Index (HQI) framework for high-throughput workload-aware batch processing of hybrid queries. Our contributions target two dimensions: (i) workload-awareness in vector index design (workload-aware vs. workload-oblivious); and (ii) query processing setting (batch vs. online). The space created by adopting optimizations that target these two dimensions is shown in Figure 1. In the same figure, we also place KG tasks that can benefit from optimizations in the corresponding quadrant. Current vector databases adopt index designs that limit themselves to one quadrant: the vector index is workload-oblivious and queries are processed online. In contrast, our KG-related workloads allow us to explore the remaining space. For example, a related queries or entities service can leverage workload-aware search (*e.g.*, online related KG queries with *a priori* workload) to minimize query execution time. Similarly, both this service and link prediction (without *a priori* workload) can benefit from batch-optimized execution. The two optimization opportunities are complementary. To this end, HQI proposes a suite of solutions including a *workload-aware vector index* and a *multi-query optimization technique*. These solutions cover the optimization space of interest. Moroever, we show that these optimizations can significantly improve query processing throughput compared to baseline solutions employing a widely-used open-source library for vector similarity search [23]. Our contributions are:

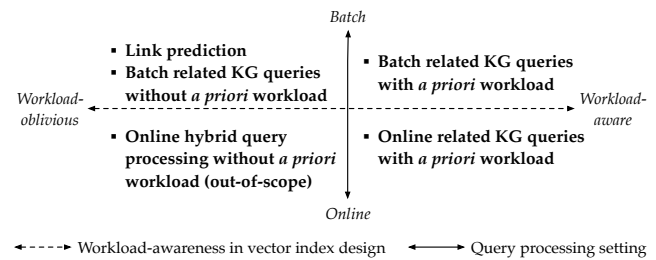

Figure 1: Optimization space for vector search and assignment of supporting KG tasks.

**Workload-aware vector index (Section 4):** First, we introduce a new workload-aware index for vector databases. Specialized vector indexes that either partition the data or form multi-level indexes over centroids are commonly used in vector databases to speed-up vector similarity search [12, 42, 45]. Here, we use past workload information to guide the partitioning of the vectors in the underlying index in a way that hybrid queries can be answered by accessing as few partitions as possible. Our approach is motivated by the fact that the relational predicates in our target hybrid query workloads exhibit *filter commonality* and *filter stability* [38], two properties that enable data layout optimizations tailored to the characteristics of a given workload. To this end, we extend the concept of *query-data routing trees (qd-trees)* [46] to consider both vectors and relational predicates from a hybrid query workload when generating physical data layout at data loading time. In addition, we discuss how this data layout enables joint evaluation of attribute and similarity predicates during query runtime. When used as a standalone optimization, our proposed workload-aware vector indexing scheme significantly reduces the number of tuples scanned to answer hybrid queries, resulting in 4× improvement in throughput for an online related KG query use case compared to workload-oblivious approaches.

**Batch query optimization (Section 5):** Second, we introduce a multi-query optimization technique that (i) batches queries with similar attribute and vector similarity constraints; and (ii) performs batch vector distance computation against a posting list of vectors obtained from a clustering-based index over the vectors. This optimization is motivated by the fact that our target hybrid query workloads need to be evaluated in a batch setting, enabling computation sharing across queries. This optimization is applicable to any clustering-based vector index. When compared to executing queries one-by-one using a system optimized for online search, our approach provides 17× and 19× improvement in batch query processing throughput for batch related KG queries (without considering prior workloads) and link prediction, respectively. Furthermore, combining the query batching technique with the workload-aware index layout yields up to 31× thoughput improvement for the related KG queries use case compared to prior approaches.

Table 1: Query workload characteristics. Rows are sorted from the lowest to highest selectivity.

| Template | Historical Snapshots (equal time windows) | | | | Feasible KG entities (%) |
|---|---|---|---|---|---|
| | $t_0$ | $t_1$ | $t_2$ | $t_3$ | |
| T1 | 15% | 17% | 17% | 18% | < 0.005% |
| T2 | 26% | 26% | 26% | 26% | <0.1% |
| T3 | <1% | <1% | <1% | <1% | <0.1% |
| T4 | 24% | 20% | 20% | 20% | <0.5% |
| T5 | 11% | 12% | 11% | 12% | <0.5% |
| T6 | 2% | 2% | 2% | 2% | <1% |
| T7 | 3% | 3% | 4% | 3% | 2.5% |
| T8 | 15% | 15% | 15% | 14% | 30% |
| T9 | <1% | <1% | <1% | <1% | 58% |
| T10 | 4% | 4% | 4% | 4% | 60% |

Finally, we provide several micro-benchmark experiments using industrial KG workloads of hybrid queries and synthetic benchmarks and conclude with a discussion of findings and recommendations for new optimizations in vector database systems (Section 6).

## 2 PRELIMINARIES

We discuss preliminary concepts that are necessary in the remainder of the paper. First, we describe the requirements of industrial KG workloads that motivate our study and highlight their key characteristics (Section 2.1). Then, we review existing solutions for processing hybrid queries (Section 2.2), and discuss their limitations with respect to the workloads we consider (Section 2.3).

### 2.1 Workload Characteristics

In Section 1, we discussed how hybrid vector search is used in Saga to power several applications such as finding related entities, link prediction, and detecting erroneous facts [19]. We use the related KG queries use case as a running example to introduce necessary concepts. Nonetheless, all use cases exhibit the same or subset of the characteristics of this use case and all can benefit from the optimizations in this work.

The related KG queries use case requires analyzing historical queries over a KG to either suggest new *related* queries to past queries or identify entities related to input KG queries (see motivating examples in Section 1). As discussed in our prior work on Saga [19], one can use technologies that embed a KG into a vector space [30, 40], *i.e.*, every entity is associated with a vector representation. We also define *relatedness* between two entities in a KG to be the similarity between their vector representations. Given the above, both tasks correspond to evaluating a collection of *hybrid queries* over a KG, *i.e.*, queries where we want to find KG entities that satisfy a *conjunction of relational constraints* and their vector representations are similar to the vectors associated with the entities given as input to historical KG queries.

*Example.* Following the example in Section 1, the user query Q, "How tall is Taylor Swift?" requires a query corresponding to a predicate template "Height" over the entity "Taylor Swift". Finding related queries to this user query requires performing hybrid search to identify entities whose vector is similar to that of "Taylor Swift" and that have a non-NULL "height" value and are type of "Person".

To power the related KG queries service, we construct a hybrid query workload by using past KG queries. The corresponding distribution of predicate templates and KG entities that satisfy the relational predicates for these templates is determined by the past KG queries. For the constructed hybrid query workload, we use the term *selectivity* to refer to the probability that an entity satisfies a predicate, *i.e.*, the lower the selectivity the smaller the number of entities satisfying the predicate will be [37]. Next, we analyze four historical snapshots of randomly sampled and aggregated industrial query workload each spanning equal time windows. The sampling was done in a way to not alter the distribution of predicate templates within each snapshot. Table 1 summarizes distributional characteristics for the top-10 predicate templates in these historical snapshots. Our key observations are:

- The distribution of predicate templates exhibits *filter commonality* [38], *i.e.*, a small number of predicates are commonly used by the majority of queries. For instance, ≈80% of queries in the workload corresponds to only four templates.
- Predicate templates are repeatedly used over time and their distribution exhibit *filter stability* [38], *i.e.*, the majority of predicate templates in queries at a given time have already occured in the past. Furthermore, the composition of query workload does not drastically change across snapshots, *i.e.*, the ratio of a particular template does not change across snapshots in Table 1.
- Hybrid queries exhibit a variety of predicate templates that consist of complex attribute constraints beyond simple numerical comparisons. For instance, some commonly occurring predicate templates in this workload consist of set membership and non-NULL checks on KG entity attributes.
- Predicate templates exhibit a wide range of filtering capabilities. For instance, less than 0.005% KG entities satisfy the predicate template with the lowest selectivity (T1). Meanwhile, more than half of the KG entities satisfy the predicate template with the highest selectivity (T10).
- The set of attributes an entity has is impacted by its type, resulting in a varying distribution of attribute occurrences and correlations across entity types.

These characteristics necessitate workload-aware techniques to ensure consistently high-throughput across workloads processed over time. Especially, due to the filter commonality and filter stability properties, we can benefit from building a workload-aware index once and reuse it over time.

### 2.2 Methods for Evaluating Hybrid Queries

An emerging class of data management systems provide query processing primitives to support hybrid queries [1–3, 25, 42, 44]. Given a hybrid query, these systems facilitate searching for objects in a vector database where (i) the vector representation of the object is similar to the query vector, and (ii) the object attributes satisfy the structured attribute constraint. Existing systems commonly process these two sub-tasks disjointly and employ one or more of the following evaluation strategies:

**Strategy A – Exhaustive search:** This approach constructs a traditional relational index (*e.g.*, B-tree index) over attributes and evaluates the attribute constraint using an index scan. Then, it exhaustively computes the distances between the query vector and all vectors that satisfy the attribute constraint to perform similarity search. This strategy produces exact results with respect to both the vector similarity and attribute constraints.

**Strategy B – Attribute filtering then vector search:** This strategy relies on two disjoint indices over the database: (i) an approximate nearest neighbor (ANN) index over the vectors [20, 21, 23, 28]; and (ii) a relational index over the vector attributes. Similar to Strategy A, this approach first evaluates the attribute constraint to obtain a candidate set. Then, it uses the ANN index to perform similarity search over the resultant vectors. This strategy is commonly implemented by generating a bitmap from the identifiers (IDs) of the candidate vectors to filter out vectors that do not satisfy the attribute constraint during the index traversal.

**Strategy C – Attribute-based partitioning:** This approach generates a two-tier vector index by range partitioning the vectors based on a frequently searched attribute and constructing an ANN index within each partition. Given a hybrid query with a predicate over the partitioning attribute, this approach first identifies the partitions whose assigned range satisfies the query constraint, then performs either exhaustive search or ANN search within each qualifying partition depending on the partition size.

**Strategy D – Vector search with post-filtering:** This strategy constructs an ANN index over the vectors, and first performs a vector similarity search using the ANN index, then it filters out candidates that do not satisfy the attribute constraint.

Strategy A is an exhaustive solution that produces exact results with respect to the vector similarity search constraint. The performance of Strategy A is determined by the selectivity of the attribute constraints and consequently this approach is preferred either for small datasets or hybrid queries with highly selective attribute constraints. Strategies B, C, and D aim to address these scalability limitations over large-scale datasets by augmenting a specialized index for vector similarity search and are commonly adopted in industrial solutions for hybrid query processing [1–3, 25, 42, 44]. All the above approaches are designed for online query evaluation, and as we discuss next, they do not fit the requirements for high-throughput batch evaluation of hybrid query workloads with characteristics as those outlined in Section 2.1.

## 2.3 Limitations of Existing Methods

**No optimizations for batch processing:** Existing approaches for hybrid query processing are optimized for online query execution. Although these systems can functionally support batch workloads by processing each query individually, they cannot utilize the data layout and query processing optimizations that are possible in the batch setting. Also, hybrid query workloads we target exhibit commonalities between the query vectors as well as between the relational predicates over the attributes, providing opportunities for executing similar queries together for efficient index traversals, which is not considered by the online query processing solutions. In Section 6, we empirically show that batch query optimization by itself provides up tp 17× improvement in execution time for the related KG queries workload compared to using online solutions.

**Disjoint execution of sub-queries:** Existing systems commonly treat attribute filtering and vector search as two separate sub-queries whose results are merged to generate final results. They either perform attribute filtering and vector search separately over the corresponding attribute and ANN indices, and combine the results (Strategies A, B, and C); or perform attribute filtering as a post-processing step after ANN search (Strategy D). The first approach leads to unnecessary computational overhead since the pruning power of two sub-queries cannot be combined; the vector search performs redundant vector similarity computations for tuples that the attribute filter would otherwise prune. The latter approach suffers from low recall since ANN search is oblivious to the vector attributes and performing attribute filtering as a post-processing step might prune a large portion of the ANN search results. Jointly considering vector search and attribute filtering provides opportunities for efficient hybrid query evaluation by eliminating redundant computations without sacrificing result quality.

**Limited support for attributes:** Existing strategies provide limited support for attribute constraints, *e.g.*, attribute constraints are limited to numerical comparisons or exact text match [25, 42, 44]. A common approach for structured constraints over attributes is to use a range partitioning of vectors over a frequently used numerical attribute (Strategy C). Although this approach is effective when queries can be routed to a single partition, queries with constraints over a non-partitioning attribute must be evaluated over all partitions. The storage overhead of supporting multiple attributes grows linearly with the number of attributes as it involves replicating and partitioning the vectors for each attribute.

**Lack of workload awareness:** Optimized for online query execution, existing solutions do not utilize the available workload information. Structured predicates in attribute constraints and the vectors representing real-world entities are often correlated in real-world hybrid query workloads. Consider the vector representation for the song "Billie Jean" with the entity type "Song". It is likely to be similar to vectors representing other songs than to vectors representing cities. These correlations, when considered, can significantly improve the batch query execution performance by informing the index design as we show in Section 6.

## 3 BATCH PROCESSING OF HYBRID QUERIES

### 3.1 Problem Definition

DEFINITION 1 (VECTOR DATABASE). *$V$ represents the set of tuples (vectors) in the form $t = (id, e, a)$ where $t.id$ is the tuple identifier, $t.e$ represents a $d$-dimensional real-value vector, and $t.a$ represents the attribute values associated with that tuple as discussed next.*

*$A$ represents a finite set of attributes where the domain of each attribute $A_i$ is $D_i$. A tuple can be associated with a subset of attributes from $A$; w.l.o.g., we represent $t.a$, the attributes of a tuple $t$ as a $|A|$-dimensional vector where $t.a[i] \in D_i$ represent the value of attribute $A_i$ for tuple $t$ (with null values set for non-existent attributes).*

We neither make any assumption about how the vectors and their attributes are generated nor about their sizes. Devising an embedding method with properly tuned parameters for a task such as related entities search is an orthogonal topic and is outside the scope of this work; a host of different models can be used [11, 14, 16, 32, 35, 39]. We also assume $V$ to remain static over the course of batch query processing. Updates to $V$ are accumulated over a longer time period and all indices used for batch query processing are rebuilt to reflect the updates.

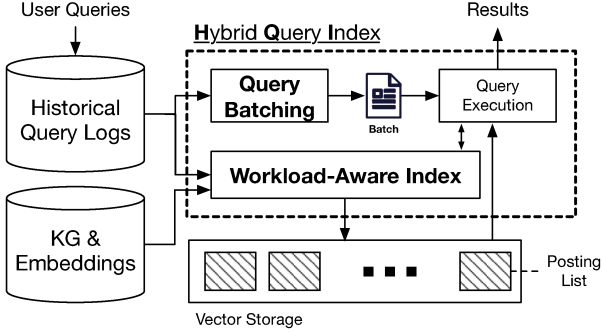

Figure 2: An overview of HQI for batch processing of hybrid vector similarity search queries.

DEFINITION 2 (HYBRID QUERY (HVQ) WORKLOAD). *$Q$ represents a workload of hybrid queries where each hybrid query $q \in Q$ contains the following: (i) a feature vector $e$, and (ii) an attribute constraint $f$ as a conjunctive Boolean predicate, i.e., $f : p_1 \wedge \cdots \wedge p_k$. W.l.o.g., we assume that a predicate $p \in f$ is either a unary comparison ($A_i \otimes x$), a set membership check ($A_i$ IN $\{x_1, \cdots, x_j\}$) or an existence check ($A_i$ IS NOT NULL) where $\otimes \in \{<, \leq, >, \geq, =\}$ is a binary predicate, $A_i \in A$ is an attribute, and $x_j \in D_i$ is a value from $A_i$'s domain.*

DEFINITION 3 (BATCH HVQ WORKLOAD PROCESSING). *Given a vector database $V$ and a hybrid query workload $Q$, the task of batch processing of a hybrid query workload can be modeled as:*

```
SELECT Q.e, Q.f, V.id as candidate FROM Q, V
 WHERE IsFilterValid(Q.f, V.a)
 ORDER BY Related(Q.e, V.e) LIMIT K
```

IsFilterValid *checks if the attributes of a tuple satisfy the constraints, i.e., $\bigwedge_{p_i \in Q.f} (p_i(V.a))$ evaluates to* True, *and* Related *computes the vector similarity between the query vector $Q.e$ and a database vector $V.e$.*

We can use the above formulation to express the example from Section 2.1, *i.e.*, finding top-K related entities to "Taylor Swift" which also satisfy the "Height" predicate template, as follows:

```
SELECT V.id as candidate FROM V
 WHERE 'Person' IN V.a['type'] AND V.a['height'] IS NOT NULL
 ORDER BY Related(Q.e, V.e) LIMIT K
```

Here, $Q.e$ is the embedding vector for the KG entity corresponding to "Taylor Swift".

## 3.2 Solution Overview

Based on the observation that scanning tuples and computing similarities between their vectors is the main bottleneck for hybrid query execution, we aim to eliminate unnecessary tuple scans and vector similarity computations by limiting the search space. For this purpose, we introduce HQI, a hybrid query processing system consisting of: (i) a workload-aware index layout that utilizes the knowledge of past queries for eliminating unnecessary tuple scans; and (ii) a batching technique that maximizes computation sharing across a batch of queries. A recent trend in database systems is to custom fit system components to a given workload and data, in this way providing significant performance improvements over their traditional counterparts in the context of relational workloads [4, 7, 17, 38, 46]. We build on these advances and introduce workload-aware optimizations within HQI, which can benefit use cases exhibiting stable distribution of predicate templates over time such as the related KG query search use case.

HQI takes a vector database and a sample of a historical workload of hybrid queries as input and builds a vector index with a custom layout that is optimized to perform well for the given workload. HQI utilizes the distribution of vectors and relational predicates in the query workload to inform the index layout, and executes queries in batches based on the commonality of vector similarity and attribute constraints. HQI's workload-aware vector indexing scheme can be employed standalone to benefit hybrid query processing even in the online setting by reducing unnecessary tuple scans. Similarly, the standalone application of the query batching technique in HQI can benefit applications that require high-throughput evaluation of hybrid queries in the batch setting by pruning redundant vector similarity computations, regardless of the availability of *a priori* knowledge about queries. Figure 2 shows a birds-eye view of our solution. Two key components of our solution are:

**Workload-aware vector index:** We introduce a vector partitioning scheme that uses the distribution of the attributes associated with vectors, the attribute constraints, and similarity of vectors present in the hybrid query workload. Our proposed scheme produces a concise description of all tuples within each resulting partition, called *semantic description*, that helps decide if the tuples within that partition can answer a given hybrid query. We then use the resulting partitioning scheme to generate an index layout that enables us to process a batch workload of hybrid queries by accessing vectors from as few partitions as possible. For workloads where templates are stable over time, *i.e.*, only a fraction of templates are newly introduced over time, such as the industrial workload presented in Section 2.1, the resulting index can benefit unseen queries without the need for re-indexing.

**Batch query execution:** When executing a batch of hybrid queries using these partitions, we first batch together all queries with the same attribute constraint (*e.g.*, predicate templates in the related KG queries workload) and route those queries to the appropriate partitions based on the partitions' semantic descriptions. After routing queries to the relevant partitions, we perform a batch vector distance computation against the vectors within that partition using a single matrix multiplication. We then merge results from across partitions.

## 4 WORKLOAD-AWARE VECTOR INDEX

Existing indexes for vector similarity search commonly rely on partitioning to scale search for input queries. A common strategy is to group the database vectors into clusters, each represented by a centroid, and use these clusters to partition the underlying vector database. At query time, only the partitions whose corresponding centroids are the closest to the query vector are scanned.

When vectors in a database are associated with relational attributes, existing solutions for processing hybrid queries combine attribute-based partitioning with vector partitioning methods (Strategy C described in Section 2.2). Such partitioning typically considers *only one attribute* — typically a commonly queried attribute — and uses *range partitioning* to first split the database into a set of attribute-based partitions. Then, a separate vector index such as *Inverted File Index (IVF)* [21] or *Hierarchical Navigable Small Worlds (HNSW)* [28] is constructed within each partition. Given such a partitioning, a hybrid query with an attribute constraint over the partitioning attribute can be evaluated by considering only the partitions that satisfy that predicate, effectively limiting the scope of vector similarity search by pruning partitions that do not satisfy the predicate. While this approach can be effective in use cases where online queries conform to a specific relational attribute, it does not satisfy the requirements of workloads we target in this work; recall that we aim to process hybrid query workloads that feature attribute constraints with predicates over multiple attributes (Section 2.1). The above partitioning scheme cannot utilize the pruning power of attribute constraints when queries impose predicates on non-partitioning attributes.

Designing an index for processing hybrid queries with high pruning capabilities in the presence of multiple attributes involves answering the following technical questions:

- How to custom fit the partitioning and data layout to the dataset and the workload?
- How to minimize the size of the index by avoiding replication of database vectors across partitions over multiple attributes?
- How to incorporate vector similarity and complex attribute constraints with multiple attributes in a unified manner?

The following section introduces a workload-aware vector index design that addresses these questions.

### 4.1 Workload-aware Partitioning and Indexing

**Algorithm 1:** ConstructBalancedQDTree

**Input:** $P$ = partition and its vectors; $Q$ = queries;
1 **function** ConstructBalancedQDTree($P, Q$)
2    $C \leftarrow$ ExtractCutPredicates(Q)
3    root $\leftarrow$ init_qdtree_node()
4    **if** $|P|$ > MIN_SIZE **then**
5        split_predicates $\leftarrow \emptyset$
6        left_split_size $\leftarrow 0$
7        **while** *left_split_size* $\leq |P| / 2$ **do**
8            predicate $\leftarrow$ GetMinCostPredicate($P, Q$)
9            $C \leftarrow C \setminus$ {predicate}
10           split_predicates $\leftarrow$ split_predicates $\cup$ {predicate}
11           $(P_L, P_R) \leftarrow P$.split(split_predicates)
12           left_split_size $\leftarrow |P_L|$
13       $(Q_{left}, Q_{right}) \leftarrow Q$.split($P_L.B, P_R.B$)
14       root.left $\leftarrow$ ConstructBalancedQDTree($P_L, Q_L$)
15       root.right $\leftarrow$ ConstructBalancedQDTree($P_R, Q_R$)
16   **return** root

The goal of our workload-aware vector index design is to generate a partitioning of a vector database for a given hybrid query workload where attribute constraints of queries feature various predicates over multiple attributes. To obtain such an index, we extend the *qd*-tree data structure [46], which is a generalization of the classical kd-tree [8] for workload-aware partitioning of relational

**Algorithm 2:** GetMinCostPredicate

**Input:** $P$ = partition and its vectors; $Q$ = queries;
1 **function** GetMinCostPredicate($V, Q$)
2    $C \leftarrow$ ExtractCutPredicates(Q)
3    (min_predicate, min_cost) $\leftarrow$ (nil, 2|Q|)
4    **for** *predicate* $\in C$ **do**
5        $(P_L, P_R) \leftarrow P$.split(*predicate*)
6        $(Q_L, Q_R) \leftarrow Q$.split($P_L.B, P_R.B$)
7        cost $\leftarrow Q_L$.size() + $Q_R$.size()
8        **if** *cost* < *min_cost* **then**
9            (min_predicate, min_cost) $\leftarrow$ (predicate, cost)
10   **return** min_predicate

tables. In a nutshell, a qd-tree creates a tree-based partitioning of a table by iteratively splitting the tuples using predicates present in a given query workload. The root node of a qd-tree corresponds to the entire dataset and is associated with a Boolean cut predicate $p$ such that its left subtree consists of tuples that satisfy $p$ and its right subtree consists of tuples that satisfy $\neg p$. The cut decisions are made such that the number of partitions that need to be accessed for processing the given query workload is minimized. Given a qd-tree, a disjoint partitioning of the dataset can be generated from its leaves by routing each tuple from the root to the leaves by evaluating the nodes' predicates on the tuple. A concise representation of tuples within each partition, called *semantic description*, is generated from its ancestor' cut predicates and used to decide if the tuples within that partition can answer a given query.

The qd-tree's ability to consider various predicates over multiple attributes from a given workload enables us to address the first two challenges. The key issue for adapting qd-trees to hybrid query workloads is to consider both vector similarity and attribute constraints on the tuples and queries when making cut decisions. To accomplish this, we design a transformation methodology to represent vector similarity constraints using categorical attributes (Section 4.1.1). This enables us to represent both vector similarity constraints and relational predicates on attributes in a uniform manner. Then, our modified qd-tree construction algorithm generates a partitioning of vectors by jointly considering vector search predicates and structured attribute predicates across multiple attributes (Section 4.1.2). The result of this approach is a partitioned index design that accounts for both the underlying vector database and the predicates of the given hybrid query workload. Finally, we process a batch of queries by routing them to the relevant partitions based on the semantic description of the partitions (Section 4.1.3).

*4.1.1 Incorporating vector similarity constraints.* A qd-tree relies on unary Boolean predicates (range or equality predicates) extracted from a query workload for making cut decisions, and for data and query routing. We augment the query vectors and the database tuples with a categorical attribute based on vector similarity constraints to incorporate vector similarity constraints into qd-tree construction. First, we apply k-means clustering to the vectors in $V$ to obtain a list of centroids $C$, and assign an integer identifier $c.id$ to each centroid *s.t.* $c.id \in [0, |C| - 1]$. Then, each tuple in $V$ is mapped to the centroid with the highest similarity, *i.e.*, nearest centroid in vector space, and the centroid identifier is used as a categorical attribute representing the vector. Similarly, the vector similarity constraint in each query $q \in Q$ can be transformed to a Boolean predicate by representing the vector with their closest $m$ centroids. Intuitively, this gives us a subset of tuples from $V$ that are relevant to a given query based on their vector similarity where the parameter $m$ controls the size of this subset. After the transformation, each tuple $t \in V$ has a centroid assignment $t.c$ and each query has a set of $m$ centroid assignment $q.c = \{c_0, c_1, \ldots c_m\}$. For the rest of the paper, we use $t.c$ and $q.c$ to denote the closest centroid for a tuple $t$, and the set of $m$ closest centroids for a query $q$, respectively. Figure 3a and Figure 3b illustrates centroid assignments for $V$ and $Q$, respectively, where $|C| = 2$ and $m = 1$.

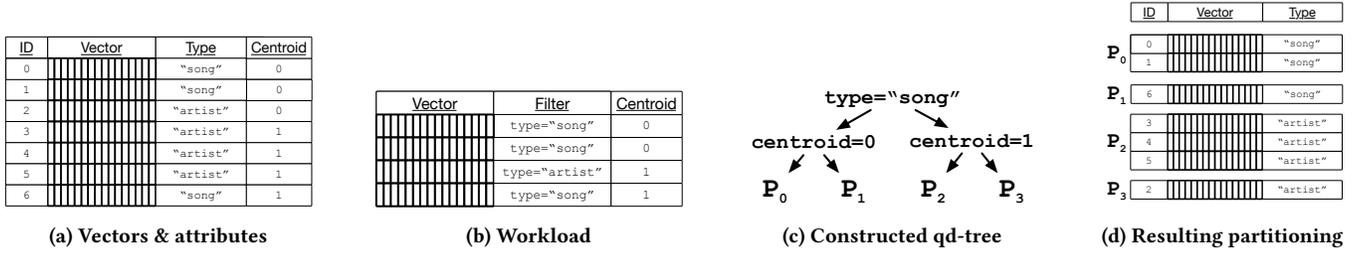

(a) Vectors & attributes    (b) Workload    (c) Constructed qd-tree    (d) Resulting partitioning

Figure 3: (a) A sample vector database, (b) hybrid query workload, (c) and example qd-tree constructed over vectors in (a) based on the workload (b), and (d) 4-way partitioning of tuples in (a) based on the qd-tree in (c).

*4.1.2 Partition generation.* Given a hybrid query workload, we extract all unary Boolean predicates that appear in attribute constraints, and we extract the centroids from vector similarity constraints using the transformation strategy described in Section 4.1.1. For instance, the set of cut predicates extracted from the example workload in Figure 3b are as follows:

$$\{(t.A_{type} = \text{song}), (t.A_{type} = \text{artist}), (t.c \in \{c_0\}), (t.c \in \{c_1\})\}$$

Each node in a qd-tree corresponds to a partition of tuples $V$ and is associated with a semantic description $B$ that is used to process cuts. As in [46], the semantic description of a qd-tree node is a bitmap denoting which of the extracted predicates evaluate to true for any tuple in the partition.

Optimal qd-tree construction is a computationally hard problem [46], therefore, we adopt a greedy construction process similar to the one presented by Yang et al. [46]. We observe that the default greedy strategy presented by Yang et al. [46] results in *imbalanced qd-trees* with high construction time and low pruning power, *i.e.*, the number of partitions that can be skipped for a given hybrid query workload is low. This is due to the presence of highly selective predicates from attribute constraints (as shown in Table 1) and centroid assignments. To this end, we propose a modified greedy qd-tree construction algorithm (ConstructBalancedQDTree procedure presented in Algorithm 1) to create balanced qd-trees. Algorithm 1 starts with a single partition that contains all tuples in $V$, and it recursively creates balanced splits of the nodes until a minimum partition size, MIN_SIZE, is reached (line 4). To ensure splits are balanced, our algorithm chooses multiple predicates from the set of cut predicates for splitting the current subtree root instead of a single one. For each split decision, Algorithm 1 iterates over the set of predicates and greedily selects the predicate minimizing the splitting cost (line 7 – 12). Recall from Section 3.1, our objective is to minimize the number of tuple scans for evaluating $Q$. Consequently, the greedy predicate selection aims to minimize the number of partitions that need to be accessed when evaluating a hybrid query workload $Q$ (*i.e.*, maximize the number of partitions that can be skipped for evaluating $Q$). Given a partitioning $P = \{P_i, \cdots P_n\}$ of $V$ and a workload $Q$, $C(P, Q)$ computes the total number of partitions that need to be accessed as follows:

$$C(P, Q) = \sum_{P_i \in P} |P_i| \sum_{q \in Q} \begin{cases} 1, & Subsumes(P_i.B, q) \\ 0, & otherwise \end{cases} \quad (1)$$

Here, $|P_i|$ corresponds to the number of tuples in a partition $P_i$, $P_i.B$ corresponds to the semantic description of a partition $P_i$, and Subsumes is a binary function indicating whether partition $P_i$ needs to be accessed when processing $q$, *i.e.*, $P_i$'s semantic description $P_i.B$ subsumes $q$'s constraints. Algorithm 1 invokes GetMinCostPredicate procedure described in Algorithm 2 to minimize the above cost function. For each predicate in a workload $Q$, the splitting cost in Algorithm 2 is computed as the number of queries that need to be routed to both subtrees, *i.e.*, the number that need to access both partitions after the split (lines 5 – 7). Figure 3 shows an example qd-tree and database partitioning constructed over a toy database and workload.

*4.1.3 Query processing.* Given a qd-tree constructed by Algorithm 1, its leaf nodes define a disjoint, complete partitioning of the original database. The partitions' semantic descriptions are used for determining the set of partitions that need to be accessed for evaluating a given query. An incoming query is routed to all partitions whose semantic description subsumes the attribute constraints and centroid assignments of that query. Note that this qd-tree-based partitioning of a database is orthogonal to vector query processing, and a separate ANN index can be created within each partition similar to the Strategy C described in Section 2.2. In this work, we adopt a per-partition clustering-based ANN index, *i.e.*, we create a separate IVF index within each partition and perform vector similarity search over the IVF indices of the partitions whose semantic description subsumes the query predicates.

*4.1.4 Analysis.* HQI construction consists of two tasks: (i) building a qd-tree using Algorithm 1; and (ii) training an IVF index for each partition corresponding to the qd-tree leaves. In the worst-case, if the resulting qd-tree becomes degenerate, the complexity of Algorithm 1 grows as $O(|V|^2)$. This can only happen when the workload consists of $|V|$ many cut predicates and each predicate evaluates to true for a unique set of $|V|-1$ rows in $V$. However, this is a pathological example and we do not observe such imbalance in qd-trees generated by Algorithm 1 in practice. In the case of a balanced qd-tree, the complexity of Algorithm 1 grows as $O(|V| \log p)$ where $p$ is the number of vector partitions in the resulting qd-tree. For each such partition $P_i$, training an IVF index with $\sqrt{|P_i|}$ centroids has a time complexity of $O(|P_i|\sqrt{|P_i|})$. For a balanced qd-tree, we have $\forall_i : |P_i| = \frac{|V|}{p}$. Therefore, the total cost of building $p$ instances of IVF indexes is $O(|V|\sqrt{\frac{|V|}{p}})$. Thus, the overall time complexity of building a HQI for a balanced qd-tree is $O(|V|(\log p + \sqrt{\frac{|V|}{p}}))$.

## 4.2 Attribute constraint aware ANN index

The workload-aware indexing strategy described in Section 4.1 reduces the total number of tuples scanned by pruning partitions irrelevant to the query. Within each partition, we employ a clustering-based ANN index for performing vector similarity search. We can eliminate unnecessary vector distance computations within the ANN index by constructing a structured index over attributes of tuples in $V$ and using sideways information passing between the two indices, *i.e.*, by pushing a succinct representation of attribute constraints into the ANN index. We achieve this by evaluating the attribute constraint on a structured index (such as *B-Tree Scan*) and construct a bitmap that encodes the set of resultant tuple identifiers. We then pass this bitmap encoding of the filtered tuple identifiers to the ANN index along with the query vector. We modify the corresponding search algorithm of the underlying ANN index such that this bitmap is used to skip distance computations for vectors that would be discarded anyway due to not satisfying the attribute constraints. Specifically, we modify the index scan phase of clustering-based indexes such as IVF to use bitmap tests to skip vector distance computations while scanning posting lists. A similar strategy can be applied to graph-based indices such as HNSW by using bitmap tests to exclude vectors from the candidate set during the graph traversal phase [12, 44].

Pushing bitmaps into the ANN index to prune unnecessary vector distance computations is orthogonal to the partitioning scheme and can be used in a standalone manner to incorporate attribute constraints into existing vector similarity search techniques. Indeed, this is similar to Strategy B (Section 2.2) and is gaining popularity for the online evaluation of hybrid queries in open-source vector similarity search libraries and commercial systems [23, 42, 44]. One potential limitation with its standalone application is that vector index construction is oblivious to the structured attributes. Therefore, evaluating queries with highly selective constraints might require accessing a large portion of the index. Combined with our workload-aware index layout, this approach can reduce both the portion of the index that needs to be accessed and the number of distance computations performed, as shown in Section 6.

## 5 BATCH QUERY OPTIMIZATION

The workload-aware index design and query processing strategy from Section 4 utilize the information of past queries and reduce the total number of tuple scans and vector distance computations by skipping partitions irrelevant to a given query. The batch nature of query processing provides optimization opportunities for further eliminating redundant computations that are orthogonal to workload indexing. Specifically, we observe from our motivating workload that a subset of query vectors and attribute constraints

**Algorithm 3:** BatchProcessing

**Input:** $I$ = IVF index; $Q$ = query workload; $k$ = Number of nearest neighbors to return; $nprobe$ = number of posting lists to scan for each query

1 **function** BatchProcessing($I, Q, k, nprobe$)
2     $C \leftarrow$ ExtractPredicates($Q$)
3     $R =$ ResultsHeap(num_results=$|Q|$, max_size=k)
4     **for** *filter* $f \in C$ **do**
5         $Q_f \leftarrow$ GroupBy($Q, f$)
6         $N \leftarrow$ FindNearestCentroids($Q_f, I, nprobe$)
7         **for** *centroid* $c \in N$ **do**
8             $Q_f^c =$ GroupBy($Q_f, c$)
9             candidates = ApplyFilter($I[c]$, filter)
10            (neighbors, dist) = KNN($Q_f^c$, candidates, $k$)
11             **for** $q \in Q_f^c$ **do**
12                 $R[q.id]$.push(neighbors[$q.id$], dist[$q.id$])
13     **return** $R$.ids, $R$.values

are used by a majority of the queries in the workload (Section 2.1). As such, given a workload of hybrid queries, we can batch queries based on the similarity of their constituent vectors and attribute constraints and execute them in batches. By sharing tuple scans and vector distance computations across a batch of queries, we can improve throughput without impacting result quality.

Batching queries by their attribute constraints is a commonly used technique in commercial solutions to amortize the cost of evaluating constraints across a batch of queries containing the same attribute constraint. This strategy is particularly effective for our motivating workload given the frequent occurrence of a small set of templates and their corresponding attribute constraints, as shown in Table 1. Batching queries based on vector similarity, however, provides an opportunity to eliminate redundant tuple scans and vector distance computations by sharing ANN index traversals, which is the main bottleneck for hybrid query execution. Consider the IVF indexes created within each qd-tree partition. A query vector is routed to a subset of posting lists based on the vector's nearest centroids, and each posting list is exhaustively scanned to find similar vectors. Our key idea is to first find the subset of queries that need to be evaluated over each posting list by grouping queries based on their vector similarity, *i.e.*, by their nearest centroids in the IVF index. Then, we calculate the distances between grouped query vectors and posting list vectors using efficient matrix multiplication instead of exhaustively scanning the posting list for each query.

We outline the aforementioned batch execution of a hybrid query workload on an IVF index in Algorithm 3. Given a hybrid query workload $Q$ and an IVF index $I$ over a set of vectors from a partition $P_i$, Algorithm 3 evaluates queries in batches as follows:

(1) Queries in $Q$ are first grouped based on the predicates in their attribute constraints (line 5).
(2) Within each group, *nprobe* nearest centroids for each query vector are obtained (line 6) and a list of query vectors per centroid is produced (line 8).
(3) For each posting list, the set of candidate vectors that satisfy the attribute constraint is obtained (line 9). Then, distances between candidate vectors and query vectors are computed as a single matrix multiplication operation (line 10).
(4) Finally, a min-heap for each query is updated with the distances computed by matrix multiplication to keep track of the $k$ nearest results for each query (line 12).

## 6 EXPERIMENT RESULTS

For our evaluation, we use both industrial and publicly available benchmark datasets and compare HQI against different recent solutions. We first describe our experimental setup (Section 6.1), then, we present an end-to-end performance evaluation of HQI (Section 6.2), present a series of microbenchmarks of our technical contributions (Section 6.3), and demonstrate the generalizability of HQI to unseen workloads (Section 6.4). The highlights of our results compared to existing solutions are:

(1) Workload-aware partitioning and indexing of the vectors provides up to a 95% reduction in the number of tuples scanned and vector distance computations, depending on selectivity.
(2) The proposed query batching strategies are effective in sharing vector distance computations across a batch of hybrid queries, providing up to an 19× improvement in overall workload evaluation time.
(3) Combined together, HQI provides up to a 31× improvement in overall workload execution time.

### 6.1 Experimental Setup

Table 2: Evaluation datasets ($n_q$ = Number of query vectors).

| Dataset | $n_q$ | Datatype | Metric | Attributes |
|---|---|---|---|---|
| SIFT-100M | 10K | 128 uint8 | L2 | synthetic |
| MSTuring-100M | 100K | 100 f32 | L2 | synthetic |
| YandexT2I-100M | 100K | 200 f32 | IP | synthetic |
| LP | - | 128 f32 | IP | entity types |
| RelatedQS | - | 128 f32 | IP | entity properties |

*Datasets.* Table 2 summarizes the properties of publicly available datasets and the randomly sampled industrial dataset we use for evaluation. The public datasets we use are SIFT-100M [34], MSTuring-100M, and YandexT2I-100M; all from the BIGANN benchmark [22]. Vectors in these datasets do not contain any attributes, and we follow prior work [42, 44] for generating synthetic attributes for vectors and query predicates. We assign each vector in the dataset two random floating point attributes, A and B. We then generate 20 range predicates, 10 on attribute A and 10 on attribute B, such that the selectivity of a range predicate $i$ over a given column is $2^{-i} : i \in [0, 9]$, i.e., $2^{-i}$ fraction of the vectors satisfy the predicate $i$. We then construct the query log by performing a Cartesian product of all 20 filters and all $n_q$ query vectors provided with the dataset resulting in $20 \cdot n_q$ queries.

For the industrial workload we use two datasets: RelatedQS and LinkPrediction (LP). For RelatedQS, we use a subset of KG entity embedding vectors and a randomly sampled and aggregated query workload from anonymized, historical queries. The hybrid query workload is obtained by transforming the historical query log into (Template, Entity) pairs, where Template is a conjunctive Boolean predicate over KG entity attributes, and Entity is the embedding vector of the corresponding KG entity. For LP, we generate the workload by sampling entites and relations from an internal KG to form (Template, Entity) pairs, however here the Template corresponds to a Boolean predicate over the type of the entity. The size of query sets accompanying these workloads range from a few millions to hundreds of millions of queries. RelatedQS represents an industrial workload that has historical queries and therefore can benefit from both workload-aware indexing and batching, while the LP workload does not have historical queries and only benefits from batching. The vector representations for KG entities are obtained by training the GraphSage model [14] over the KG.

We use all datasets in an end-to-end evaluation in Section 6.2 and use RelatedQS and MSTuring-100M for additional microbenchmark experiments in Section 6.3. For end-to-end evaluation with RelatedQS, we use all queries from one of its four equal temporal splits ($t_0$ from Table 1) for constructing the workload-aware index and use the trained index to process all queries from the same split. We run all experiments using a single AWS r5.16xlarge instance with 64 vCPUs and 512 GiB of DRAM.

*Evaluation metrics.* We use the following metrics for a given hybrid query workload: 1) total query processing time, 2) the number of tuples scanned to process the workload, 3) top-$k$ recall. We compute recall as the fraction of results present in the ground truth (obtained via exhaustive search, i.e., Strategy A from Section 2.2). Unless otherwise noted, we report runtime numbers for $Recall \geq 0.8$ for $k = 10$, where $nprobe$, the number of posting lists scanned, is tuned for each query template to reach the target recall.

*Compared approaches.* We compare our proposed techniques against the strategies described in Section 2.2. Specifically, we compare the following specific hybrid query processing strategies that are used in several commercial offerings [1–3, 25, 42, 44]:

- **PreFilter**: Strategy B using IVF index implementation based on the FAISS [23] open source library.
- **Range**: Strategy C where vectors are range partitioned based on a frequently searched attribute, and a separate FAISS-based IVF index is created within each partition.
- **PostFilter**: Strategy D implemented using FAISS where attribute constraints are evaluated after vector search.
- **HQI**: Our proposed workload-aware approach where vectors are partitioned using the qd-tree constructed using a given hybrid query workload, and queries are batched based on their attribute constraints and vector similarity.

Strategy A from Section 2.2 is omitted as exhaustively scanning all vectors is prohibitively slow for large datasets used in our experiments. Unless otherwise specified, we batch queries by their attribute constraints by default for all baselines, as it is a commonly used optimization strategy in many existing commercial solutions. Similarly, we use the bitmap pushdown technique described in Section 4.2 to reduce the number of vector distance computations for all approaches. FAISS recently introduced the capability to push attribute filters into an ANN index for pruning vector distance computations[1]. We use this feature to implement a bitmap based ID selector, which allows us to push attribute filter results into the IVF index when required. In addition, we use $\sqrt{n}$ clusters for training an IVF index, where $n$ is the number of vectors in a given partition or a dataset for a non-partitioned index. For HQI, we tune $m$ and report runtime for the best configuration ($m = 0$, where centroids are not used for query routing, is used as it provides the best end-to-end performance over RelatedQS as described in Section 6.5).

---
[1] https://github.com/facebookresearch/faiss/commit/dd814b5f146b9fb5f5f76070ff1e27b86e31a058
[2] Attribute constraints in RelatedQS and LP contain existence (IS NOT NULL) and membership (IN) checks over multiple attributes, hence, Range cannot be applied.

Table 3: Slowdown compared to HQI @ Recall >= .8

| Approach | Dataset | | | | |
|---|---|---|---|---|---|
| | RelatedQS | LP | MSTuring | SIFT100M | YandexT2I |
| HQI | 1× | 1× | 1× | 1× | 1× |
| PreFilter | 31× | 19× | 3.6× | 0.97× | 1.7× |
| PostFilter | 136× | - | 22× | 4.1× | 5.4× |
| Range | NA[2] | NA[2] | 5.22× | 1.2× | 3× |

Table 4: Index generation time compared to HQI

| Approach | Dataset | | | | |
|---|---|---|---|---|---|
| | RelatedQS | LP | MSTuring | SIFT100M | YandexT2I |
| HQI | 1× | 1× | 1× | 1× | 1× |
| PreFilter | 0.95× | 1× | 2.8× | 2.15× | 1.9× |
| Range | NA[2] | NA[2] | .85× | 0.63× | 0.58× |

## 6.2 End-to-End Performance

We first demonstrate the performance benefits of HQI by comparing its end-to-end performance to existing approaches on all datasets listed in Table 2. Note that we augment each vector and query with synthetic attributes for MSTuring, SIFT100M, and Yandex T2I as described in Section 6.1.

*Query Processing Time.* Table 3 reports the relative workload execution time for all approaches with respect to HQI. We observe that HQI provides 31× and 19× speedup in workload execution time over the best performing baseline PreFilter on the RelatedQS and LP workloads, respectively. Note that Range cannot be used for RelatedQS and LP as their attribute constraints contain existence (IS NOT NULL) and set membership (IN) predicates over multiple attributes. We observe that PostFilter is significantly slower since its vector search stage does not utilize the pruning power of attribute constraints, requiring the vector search stage to return a large number of candidates to ensure the desired recall. For LP, we were unable to acheive the target recall using PostFilter.

We observe similar trends over the public datasets, where HQI matches or exceeds the best-performing baseline. In particular, we observe that PreFilter outperforms Range as only a portion of the queries, *i.e.*, queries with predicates over the partitioning attribute can utilize the additional pruning power of Range. The performance improvements of HQI over PreFilter and PostFilter are lower over these benchmarks compared to RelatedQS because of two reasons: (i) The synthetic predicates we generate do not have the variety and skewed distribution as RelatedQS (Section 2.1), limiting the effectiveness of workload-aware indexing. (ii) SIFT100M has 10× fewer queries and both YandexT2I and SIFT100M require scanning roughly 10× fewer posting lists per query to obtain a high recall, limiting the effectiveness of batching. A detailed analysis of how much each optimization in HQI contributes to its overall performance is presented in Section 6.3. Overall, HQI provides up to an order-of-magnitude improvement in workload execution time across industrial and public ANN benchmark workloads.

*Runtime Breakdown per Query Template.* Next, we turn our attention to the runtime breakdown for each query template from RelatedQS workload in Figure 4. Recall from Table 1 that the attribute constraints present in RelatedQS contain predicates over multiple attributes with a wide range of selectivities.

We find that across a majority of the templates, HQI consistently achieves more than 10× speedup in batch query execution time compared to the best baseline (PreFilter). Runtime differences are the most significant for low- to medium- selectivity templates (T1 - T8), whereas the improvement diminishes for high-selectivity templates (T9 - T10). This is because the number of candidate tuples that satisfy these high-selectivity templates is significantly larger than that of the other templates. Consequently, the effectiveness of partitioning for pruning tuples is limited. Albeit reduced, the runtime improvement remains substantial, with a 2.5× and 17× improvement for T9 and T10, respectively.

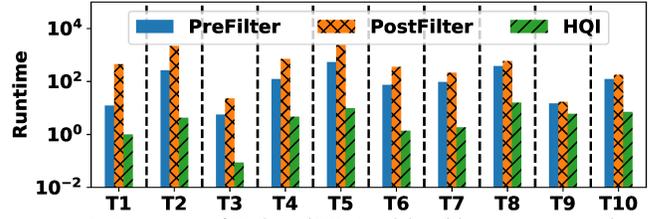

Figure 4: Runtime of RelatedQS Workload by query template. Runtime is normalized by the HQI runtime for template T1.

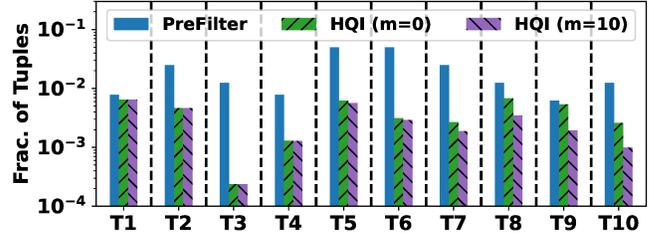

Figure 5: RelatedQS: Fraction of tuples scanned by template.

*Index Generation Time.* Table 4 shows the index generation time for each dataset normalized by that of HQI. We find that HQI's index generation time is similar to or faster than the baselines. While seemingly counter-intuitive, there are two reasons for this: (i) qd-tree construction accounts for 10% - 20% of index generation time and is not the dominant cost in index training; and (ii) training a partitioned IVF index scales better than training a single IVF index over the full dataset. The latter is due to the complexity of IVF index training: training a single IVF index over $V$ scales as $O(|V|\sqrt{|V|})$ whereas HQI scales as $O(|V|(\log p + \sqrt{\frac{|V|}{p}}))$ as described in Section 4.1.4. The faster index generation time of HQI and Range indexes on the public datasets compared to PreFilter is a direct result of training multiple IVF indexes over partitioned vectors. For LP, there is no historical query log available for indexing so the qd-tree construction is skipped, therefore the HQI index generation is identical to PreFilter. Overall, qd-tree construction does not introduce significant overheads to index generation, and HQI can provide improved generation time over unpartitioned indexes.

## 6.3 Microbenchmarks

Here we present a detailed experimental evaluation of HQI by analyzing the performance gains of our workload-aware index design (Section 4) and query batching strategies (Section 5).

*Reduction in Tuples Scanned:* We first evaluate the pruning capability of HQI against the best performing baseline – PreFilter.

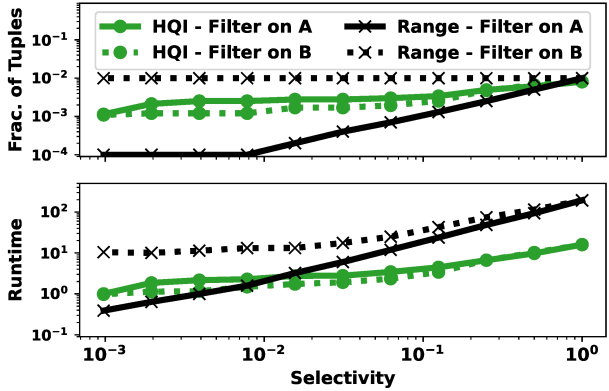

**Figure 6: MSTuring: Normalized runtime and fraction of tuples scanned by selectivity.**

We evaluate two configurations of HQI: $m = 0$ denotes that only attribute constraints are used, and $m = 10$ denotes that each query vector is mapped to the 10 closest centroids for partition generation decisions and query routing (Section 4.1.1). Figure 5 shows the fraction of tuples scanned to process a query for each template in the RelatedQS workload. Overall, we observe that a reduction in tuple scans corresponds to a reduction in runtime. HQI performs an order of magnitude fewer tuple scans than PreFilter for many templates, showing a 95% reduction for T3, in line with the runtime breakdown results reported in Figure 4. For templates with high selectivity (T9-T10), HQI ($m = 10$) provides additional pruning capability by skipping partitions that are unlikely to contain relevant vectors. However, using centroids for query routing introduces additional overhead due to finding the nearest centroids at query time, resulting in slower runtime for low- and medium- selectivity templates (T1-T7). This provides an interesting opportunity for an adaptive optimizer that controls the value of $m$ based on query selectivity, which is further discussed in Section 6.5. Overall, for the RelatedQS workload, HQI with $m = 0$ and $m = 10$ results in 77% and 80% fewer tuple scans, respectively, compared to PreFilter.

*Multi-attribute Partitioning:* We now show HQI's effectiveness for hybrid query workloads containing predicates over multiple attributes. We compare HQI against the Range baseline over the MSTuring dataset where Column A is used for range partitioning and report the total number of tuple scans and batch query execution time in Figure 6. First, we observe that the Range baseline effectively reduces the tuple scans for queries with predicates over the partitioning attribute A. However, it cannot provide any pruning when the attribute constraint of a query does not contain any predicate over the partitioning attribute (*i.e.*, attribute constraint is a predicate over the non-partitioning attribute B), increasing runtime. In contrast, HQI consistently reduces the number of tuple scans, thus the execution time, regardless of the attribute present in the predicates. As opposed to Range, which partitions over a single attribute attribute, HQI considers all predicates present in the queries and partitions over multiple attributes, resulting in runtime improvements for queries with predicates over multiple attributes.

*Query Batching:* Here we analyze the effectiveness of batch query optimization strategies from Section 5. First, we evaluate the effect of batch size on HQI and the baseline PreFilter. We then implement combinations of attribute constraint- and vector similarity-based batching over a vanilla IVF index to isolate the impact of batching from the impact of workload-aware partitioning in HQI.

We first evaluate the impact of batch size on HQI compared to the baseline PreFilter, using template T4 in the RelatedQS workload. The runtime to process a single batch of query vectors to reach a target recall of 0.8 is shown in Figure 7a. The runtime is normalized by "Batch Size = 1" case for "HQI without vector batching" scenario. Our results show that HQI can process batches over 10× faster than PreFilter for query template T4 across all batch sizes. This is due to two reasons: (i) reduction in tuples scanned per query, and (ii) smaller filtering overhead. The first reason is illustrated in Figure 5, which shows that queries using template T4 require scanning roughly 10× fewer tuples due to the pruning capability of the qd-tree. The second reason is that filters and bitmaps need only be computed within the qd-tree partitions for HQI, rather than. across the full dataset as in PreFilter. Filtering overhead is especially detrimental for small batch sizes (*i.e.*, the online setting), where it can be the dominant cost. Our experiments also show that for small batch sizes, HQI without vector batching optimization is the best option, with a crossover point at a batch size of 100. This suggests that there is an opportunity for an optimizer to determine the best execution plan based on the batch size. Although our focus is on offline workloads, our results also show that HQI benefits the online setting (*i.e.*, Batch Size = 1). In summary, HQI can process queries up to 10× faster than PreFilter across batch sizes.

We next evaluate the impact of vector similarity-based batching using the MSTuring dataset without any attributes to demonstrate the batching technique's effectiveness for vector similarity search over a vanilla IVF index. Figure 7b presents the batch query execution time for achieving different recall levels by adjusting the number of posting lists (*nprobe*) scanned for each query. We observe that the runtime remains relatively stable with our batching strategy. This is in contrast to the IVF baseline, where achieving higher recall comes at the expense of a significant increase in query execution time (due to higher *nprobe*), *e.g.*, 21× slowdown at recall 0.9. With vector similarity-based batching, distances for a group of query vectors against a posting list are computed using a single matrix multiplication instead of pairwise distance computations.

Finally, we evaluate the effectiveness of attribute constraint-based batching across queries with various selectivities (Figure 7c). Here use augment vectors and queries in MSTuring-100M with synthetic attributes as described in 6.1. First, we observe that attribute constraint-based batching is a critical optimization for performant batch processing. It provides a 300× runtime improvement over issuing queries one at a time (No batching) by amortizing the cost of evaluating attribute constraints across a group of queries. Applying vector-similarity based batching (as in Algorithm 3) provides further 18× improvement, resulting in ≈ 5000× improvement over the naive strategy of issuing queries one at a time. In summary, our batching techniques are crucial to ensure high-throughput batch processing of hybrid queries without reducing recall.

### 6.4 Robustness to Future Queries

Here, we show that the workload-aware indexing in HQI can benefit unseen queries without the re-indexing for workloads where the distibution of predicates is relatively stable over time (workloads that exhibit *filter stability* as described in Section 2.1). For this purpose, we train HQI using predicates of all the queries from only one split of RelatedQS ($t_0$ from Table 1), and measure the workload execution time for each of the four equal temporal splits of RelatedQS ($t_0 - t_3$ in Table 1) using the trained index. We then compare it against the best performing baseline, PreFilter, which does not utilize past workload information during indexing.

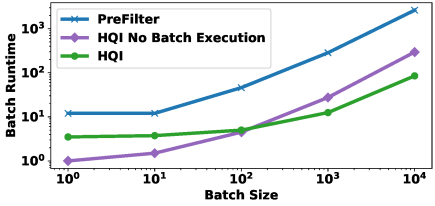
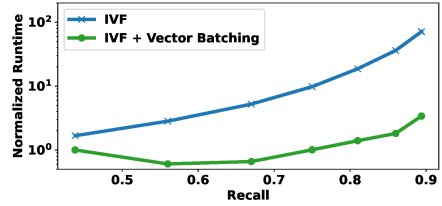
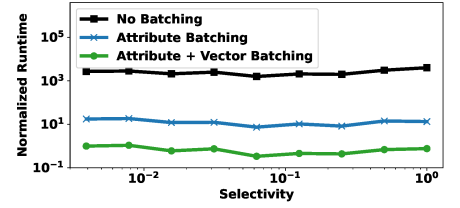

Figure 7(a): RelatedQS (Template T4): Normalized runtime to process a single batch with a varying number of query vectors in the batch.

Figure 7(b): MSTuring (no attributes): Runtime vs. Recall using an IVF index with and without vector similarity-based batching.

Figure 7(c): MSTuring: Runtime vs. Selectivity using an IVF index with and without attribute constraint-based batching.

We see from Table 5 that HQI can maintain over a 30× improvement in runtime over PreFilter across all splits. HQI can benefit unseen queries due to the relatively stable distribution of templates as shown in Table 1. As such, HQI trained over a representative query log can be re-used for future queries without the need for computationally expensive re-indexing. The small variation observed for $t_0$ through $t_3$ (≤ 5% improvement compared to $t_0$) can be attributed to the differential improvement observed for the templates (Figure 4) combined with the variations in query workload composition (Table 1) in contrast to $t_0$.

Table 5: Queries per Second (QPS) for RelatedQS workload split into four equal time intervals, where HQI is trained over queries only from $t_0$. QPS is normalized by HQI at $t_0$.

| Approach | Query Log Split | | | |
| --- | --- | --- | --- | --- |
| | $t_0$ | $t_1$ | $t_2$ | $t_3$ |
| HQI | 1× | 1.05× | 1.03× | 1.05× |
| PreFilter | .032× | .031× | .032× | .032× |

## 6.5 Discussion

Here we summarize the main findings of our evaluation and provide recommendations for new optimizations in vector database systems. First, we find that no single configuration of HQI provides the best performance across all queries and batch sizes. We observed in Figure 5 that using centroids during query routing ($m = 10$) is beneficial only to high-selectivity templates (T7-T10), while negatively affecting templates with lower selectivity due to the additional distance computations required. Furthermore, we observed in Figure 7a that there is a tradeoff point in query batch size where disabling vector-similarity based batching leads to reduced runtime for HQI. These observations show that there is an opportunity for an optimizer design that can choose the best execution policy based on the batch size and filter selectivity.

Through our experimental analysis, we find that our proposed query batching strategies are critical to achieving high-throughput query processing. When applied to the RelatedQS workload in isolation using a standard IVF index (*i.e.*, workload-oblivious setting), we obtained a 17× improvement in query processing throughput over the baseline PreFilter in Section 6.2. Furthermore, it is shown in Section 6.3 that our batch query processing technique (Algorithm 3) can provide up to three orders of magnitude runtime improvements compared to issuing queries one at a time. Despite our focus on hybrid queries, our batching strategies can be applied to traditional vector search queries. In particular, Figure 7b shows that vector similarity-based batching can still provide ≈ 20× runtime improvements compared to a standard IVF index.

Lastly, although our batching strategies are not applicable in online settings, our workload-aware partitioning strategy can benefit online hybrid query processing systems. We find that when used in isolation (*i.e.*, no batching) workload-aware partitioning and indexing (Section 4) reduces the number of tuples scanned for the RelatedQS workload by 77%, resulting in a 4× faster query processing time compared to PreFilter. This result is reinforced by Figure 7a, which shows that our indexing approach reduces the query processing time of query template T4 by up to 10× even when using a batch size of 1. Given HQI's ability to benefit unseen queries without the need for re-indexing when workload characteristics are stable over time, HQI trained using a representative sample of query logs is still effective in reducing the query processing time.

## 7 CONCLUSION

This paper describes the design and implementation of a system for batch hybrid vector similarity search query execution that exploits knowledge of the query workload to achieve high throughput. Our design is informed by the characteristics of industrial query workloads over a KG. Specifically, we introduce a workload-aware vector data partitioning scheme to fit the vector search index layout to the query workload and the vector dataset; and describe a multi-query optimization technique to reduce the overhead of vector similarity computations. Experiments using industrial query workloads and a variety of publicly available ANN benchmarks show that the combination of proposed workload-aware index design and query batching strategies provide an order of magnitude higher batch query execution throughput over existing solutions.